\documentclass[12pt]{article}
\usepackage[mathscr]{eucal}
\usepackage{bm}
\usepackage{bbold}
\usepackage{caption}
\usepackage{multirow}
\usepackage{amsmath}
\usepackage{textgreek}
\usepackage[greek,english]{babel}
\usepackage{xcolor}
\usepackage{graphicx}
\usepackage{empheq}
\usepackage{amsfonts}
\usepackage{amssymb}
\usepackage{dsfont}
\usepackage{ytableau}
\usepackage{overpic}
\usepackage{hyperref} 
\usepackage{makeidx}
\usepackage{empheq}

\newcommand{\be}{\begin{eqnarray}}
\newcommand{\ee}{\end{eqnarray}}
\newcommand{\bc}{\begin{center}}
\newcommand{\ec}{\end{center}}
\newcommand{\nn}{\nonumber \\}
\newcommand{\lb}{\label}
\newcommand{\pd}{\partial}
\newcommand{\p}[1]{(\ref{#1})}

\begin{document}

\begin{titlepage}

\vspace*{0.2cm}

\begin{center}

{\LARGE\bf Meson states  in `t Hooft model: Hamiltonian approach}

\vspace{2cm}

{\Large Andrei Smilga} \\

\vspace{0.5cm}

{\it SUBATECH, Universit\'e de
Nantes,  4 rue Alfred Kastler, BP 20722, Nantes  44307, France. }

\end{center}
\vspace{0.2cm} \vskip 0.6truecm \nopagebreak

   \begin{abstract}
We point out that the masses of the highly excited bound quark-antiquark states in  QCD$_2$ in the infinite $N$ limit may be determined in the framework of a simple  quantum-mechanical model with the potential $V(x) = \sigma |x|$. In the ultrarelativistic case,  the masses follow the pattern
$$
\mu_n^2 \ =\ \frac {g^2 N}{2\pi} n, 
$$
which coincides with the law derived in Ref. \cite{Hooft} by solving the Bethe--Salpeter equation.

In the nonrelativistic case, the levels follow the asymptotics $\epsilon_n = \mu_n - 2m  = Cn^{2/3}$, where the constant $C$ can be determined by solving the `t Hooft equation or alternatively the nonrelativistic Schr\"odinger equation.
   \end{abstract}

\end{titlepage}

\section{Introduction}

In a seminal paper  written half a century ago \cite{Hooft}, `t Hooft found the particle spectrum in 2-dimensional quantum chromodynamics with $SU(N)$ gauge group in the limit $N \to \infty$. In two dimensions, there is no gluon self-interaction (its absence is manifest if one imposes the light-cone gauge), so there are no glueballs. In the infinite $N$ limit, there are no baryons either and the only states are {\it mesons} --- bound states of a quark and an antiquark: the production of extra $q \bar q$ pairs is also suppressed. The latter fact also means that the meson states are absolutely stable in this limit. For a finite large $N$, the widths are not zero, but they are suppressed $\propto 1/N$.

't Hooft determined the spectrum by solving the Bethe--Salpeter equation for the vertices relating the bound states to their quark and antiquark constituents.

As is the case for any pioneer paper, the notations in `t Hooft's paper are somewhat sloppy. Thus, before going further, we are in a position to fix them. 

The Lagrangian of the theory reads

\be
\lb{QCD2}
{\cal L} \ =\ \frac 1{g^2} {\rm Tr} \{\hat F_{01}^2 \} + i \bar \psi \gamma^\mu (\pd_\mu - i \hat A_\mu) \psi -m \bar \psi \psi,
\ee
where $\hat A_{\mu = 0,1} = A^a_{\mu = 0,1} t^a \in su(N)$
and $\hat F_{01} = \partial_0 \hat A_1 - \partial_1 \hat A_0 - i [ \hat A_0,\hat A_1]$. Introducing the light-cone variables, $\hat A_\pm = (\hat A_0 \pm \hat A_1)/\sqrt{2}$ and similarly for $\pd_\pm$, and imposing the gauge $\hat A_- = 0$, the field density reduces to 
$\hat F_{01} = \pd_- \hat A_+$. The fermion $\psi$ (call it ``quark") belongs to the fundamental representation of $SU(N)$.

In the infinite momentum frame, the vertex of dissociation of a meson state into its constituents reduces to a simple  function $\phi(x)$, with $x \in (0,1)$ having the meaning of the fraction of the large meson momentum carried by the quark. Correspondingly, the antiquark carries the fraction $1-x$. The Bethe--Salpeter equation reduces to a comparatively simple integral equation for $\phi(x)$. It can be written in the form:\footnote{When comparing this with Ref. \cite{Hooft}, 
note that  a {\it nonstandard} definition of the coupling constant was used there:
$$
g^2_{\rm Hooft} \ =\ \frac {g^2_{\rm standard}}2.
$$}

\be
\lb{eq-Hooft}
\!\!\!\!\!\!\!\!\!\!\!\!\! \mu^2  \phi(x)  = m^2 \left(\frac 1x + \frac 1{1-x}  \right) \phi(x)  + \frac {g^2N}{2\pi} P \int_0^1 \frac {\phi(x) - \phi(y)} {(y-x)^2} dy, 
\ee
where $m$ is the quark mass and $\mu$ is the mass of the  bound state. The singularity
at $y=x$ is treated by the principal value prescription:
\be
P \int_0^1 dy\, f(y,x) \ =\ \lim_{\epsilon \to 0} \left[ \int_0^{x-\epsilon} dy \,f(y,x) + 
 \int_{x+\epsilon}^1 dy\, f(y,x)   \right]. 
\ee

  There is an infinite number of solutions for $\phi(x)$ and $\mu$. Generically, this  equation can be solved only numerically.  But the asymptotics for  high excitation levels can also be evaluated analytically. `t Hooft did it for the excitations with  masses $\mu_n$ much exceeding both relevant  scales: $\mu_n^2 \gg  m^2$ and $\mu_n^2 \gg g^2N$. In such a case,  
 \be
\lb{2D-spectrum}
\mu_n^2 \ \approx \ \frac {\pi g^2N}2 n.
 \ee
  This behavior resembles   Regge trajectories in hadron spectrum in the real 4-dimensional world, though it does not have exactly the same meaning. Regge trajectories describe an approximate  linear  dependence of hadron masses on the angular momentum in a given channel. There are many such trajectories. And in the $2D$ `t Hooft model, there is no angular momentum and all the states belong to a single trajectory where the {\it squares} of the masses depend linearly on the principal quantum number. 

Our own observation is that the law \p{2D-spectrum} may be derived in a very simple way as a solution of the relativistic quantum problem for the  potential  $V(x) = \sigma |x|$ in semiclassical approximation.  This observation is not completely original: in Refs.  \cite{Bars1,Bars2}   similar semiclassical formulas for the  meson spectrum   were derived in the string  framework.\footnote{See also  recent Ref. \cite{Kleb-WKB} where a related problem was treated in a semiclasical way.} 
 However,  no comparison with the results of Ref. \cite{Hooft}  was made in those works.

 \section{Massive particles bound  by a string}

$\bullet$ Consider first the nonrelativistic case. The Hamiltonian describing two particles of mass $m$ bound by the string is
 $$ H \ =\ \frac {p_q^2}{2m} + \frac {p_{\bar q}^2}{2m} + \sigma|x_1-x_2|. $$

 Separating out the center-of-mass motion, we may write\footnote{The reader understands of course that $x$ here has nothing to do with $x$ in Eq. \p{eq-Hooft}.}
 \be
H_{\rm cm} \ =\ \frac {p^2}m + \sigma|x|.
 \ee 
The corresponding quantum Schr\"odinger problem  can be solved exactly, the wave functions being expressed via Airy functions.
But we are not so much  interested in the exact solution. The asymptotics of the eigenstate energies at high $n$ can be determined by semiclassical reasoning. 
  
The classical momentum for a classical trajectory of energy $\epsilon$ is 
$$p^2(x) = m(\epsilon - \sigma|x|).$$
There are two turning points, $x = \pm \epsilon/\sigma$ and the  trajectory oscillates between them.
The quantization condition reads:
\be
S(\epsilon_n) \ =\ 4\int_0^{\epsilon_n/\sigma} dx  \, \sqrt{m(\epsilon_n - \sigma x)} \ =\ \frac {8\sqrt{m\epsilon_n^3}}{3\sigma} \ \approx \ 2\pi n,
\ee
giving\footnote{Cf. e.g. Eq. (16.2.38) in Ref. \cite{Shankar}.}
\be
\lb{eps-n}
\epsilon_n \ \approx \left( \frac {3\pi \sigma n}{4\sqrt{m}}  \right)^{2/3}. 
 \ee
This nonrelativistic estimate holds while $\epsilon_n \ll m$, i.e. $n \ll m^2/\sigma$.

\vspace{1mm}

$\bullet$ To find the spectrum at larger energies, one has to solve the relativistic problem. The center-of-mass Hamiltonian now reads
\be
\lb{Ham-KFG}
\hat H_{\rm cm}^{\rm rel}  \ =\ 2\sqrt{\hat p^2 + m^2} + \sigma |x|.
\ee
Again, we can evaluate the spectrum in the semiclassical approximation. The classical momentum is 
$$
p^2(x) \ =\ \frac{(E - \sigma |x|)^2}4 - m^2
 $$
 with $ E = 2m+ \epsilon$.
There are now {\it four}  turning points:
$$  \pm x_- = \pm\frac {E-2m}\sigma \qquad {\rm and} \qquad \pm x_+ = \pm \frac{E+2m}\sigma.$$

A particle may oscillate between $-x_-$ and $x_-$; quantization of this motion gives us the energy levels, but it can also tunnel through the forbidden region between
$x_-$ and $x_+$ or between
$-x_-$ and $-x_+$ and escape to infinity. This means that the levels will have a finite width.\footnote{This is an artefact of semiclassical approximation related to the fact that the Klein-Fock-Gordon equation,
$$ \left( \frac {(E - \sigma|x|)^2}4  +  \frac {\pd^2}{\pd x^2} - m^2 \right) \Psi \ =\ 0, $$
admits solutions with both positive and negative energies. If one sticks to the Hermitian Hamiltonian \p{Ham-KFG} and solves the Schr\"odinger equation {\it exactly}, the eigenvalues will stay real. \cite{Yulia}}   But let us find the real part of the energies first. The action integral is \cite{Bars2}
\be
&& S(E) \ = \ 4 \int_0^{x_-} \sqrt {\frac {E - \sigma x)^2}4 - m^2}\, dx \nn
 &&=\ \frac 1 \sigma \left[E\sqrt{E^2 - 4m^2} - 4m^2 \ln \frac {E + \sqrt{E^2 - 4m^2}}{2m}  \right]. 
 \ee

In the nonrelativistic limit, $E-2m \ll m$, this leads to the result \p{eps-n} for the energy levels. In the ultrarelativistic limit, $E \gg m$, $S(E) \approx E^2/\sigma$ and we derive 
 
\be
\lb{E-n}
E_n \ \approx \  \sqrt{2\pi \sigma n}.
 \ee

\vspace{1mm}

We go over to the widths now. The tunneling amplitude is proportional to $\exp\{-S_{\rm Eucl}\}$, where

$$
S_{\rm Eucl} \ =\ \int_{(E_n  - 2m)/\sigma}^{(E_n + 2m)/\sigma} dx\, \sqrt{m^2 - \frac {(E_n-\sigma x)^2}4}\ = \ \frac {\pi m^2}{\sigma}.
 $$ 
 The width of a state with energy $E$ is estimated as $\Gamma(E) \sim \omega e^{-2S_{\rm Eucl}}$, where $\omega \sim \sigma/E$ is the frequency with which a particle collides with the walls at $x = \pm x_-$.  We derive:
  \be
\Gamma_n \sim \frac {\sigma}{E_n} \exp \left\{- \frac {2\pi m^2}\sigma  \right\}.
 \ee
It is very small for heavy quarks, $m^2 \gg \sigma$, while if the quarks are light, the widths (appearing, as was mentioned, in semiclassical treatment of  the linear potential problem, but not in its exact solution) are of the same order as the level spacings.

\section{Comparison with QCD$_2$}

 The spectra \p{eps-n} and \p{E-n} that we derived depend on the parameter $\sigma$ of the potential --- the string tension. What is it in QCD$_2$ ? 

Consider first the Abelian case. The one-dimensional analog of the Coulomb potential created by the charge $e$ is
 \be
V^{d=1}_{\rm Coulomb} (x) \ =\ \frac {e|x|}2.
 \ee
Then $\pd^2 V^{d=1}_{\rm Coulomb} (x) / \pd x^2  \, = e\delta(x)$. The Coulomb energy of a pair
of Abelian charges separated by the distance $R$ is 
\be
E^{\rm Ab}_{\rm Coulomb} (R) \ =\ \frac {e^2 R}2.
 \ee
In the non-Abelian case, we have to replace $e^2$ by $g^2$ and multiply it by $c_F = (N^2 - 1)/(2N)  \ \to \ N/2$. We derive:
\be
\sigma_{QCD_2} \ =\ \frac {g^2 N}4.
\ee
Substituting this into \p{E-n}, we reproduce the spectrum \p{2D-spectrum}. 

An important difference  between the semiclassical KFG analysis  and QCD$_2$ is however the fact that in the latter case in the limit $N\to \infty$, the states have no width: they are absolutely stable.

The WKB asymptotics of the solution to the `t Hooft equation \p{eq-Hooft} in the nonrelativistic limit $\mu_n - 2m   \ll m$ has also been recently found \cite{Litvinov}. The result coincides with \p{eps-n} as it should. 

I am indebted to Igor Klebanov and Paul Hoyer for illuminating discussions.

\end{document}